\documentclass[showpacs,aps]{revtex4}
\usepackage{amsmath}

\usepackage{graphicx}

\usepackage{txfonts}

\begin{document}


\title{Complete analysis and generation of hyperentangled Greenberger-Horne-Zeilinger state for photons using quantum-dot spins in optical microcavities}
\author{Tie-Jun Wang$^{1}$, Si-Yu Song $^{1}$ and
   Gui Lu Long$^{1,2}$ \footnote{Email address:gllong@tsinghua.edu.cn}}
\affiliation{
$^1$ State Key Laboratory of Low-Dimensional Quantum Physics and Department of Physics, Tsinghua University, Beijing 100084, China\\
$^2$  Tsinghua National Laboratory for Information Science and
Technology, Beijing 100084, China}
\date{\today }

\begin{abstract}

We propose a scheme for the complete differentiation of 64 three-photon hyperentangled GHZ states in both polarization and spatial-mode degrees of freedoms using the quantum-dot cavity system. The three-photon hyperentangled-GHZ-state-analysis scheme can also be used to generate
3-photon hyperentangled GHZ states. This proposed hyperentangled analysis and generation device can serve as crucial components of the high-capacity, long-distance quantum communication. We use quantum swapping as an example to show the application of this device in manipulating multiparticle entanglement with polarization and spatial-mode degrees of freedoms. Using numerical calculations, we
show that the present scheme may be feasible in strong-coupling regimes with current technologies.
\\
\end{abstract}
\maketitle

\section{Introduction}

Entangled states play a critical role in quantum information processing, including quantum computation \cite{1} and quantum communication \cite{teleporation,densecoding,2}. Compared with two-partite entangled states, which are well understood, the entanglement between three or more particles is more fascinating and interesting.  As an extension from the two-qubit Bell-state to a multipartite system, the Greenberger-Horne-Zeilinger (GHZ) state \cite{ghz} is a maximally entangled state of three
or more particles, and it can be generated in different quantum systems. For example, GHZ states can be created from two pairs of entangled photons \cite{photonghz} and the excitons of coupled quantum
dots \cite{qdghz}. Researchers have also constructed GHZ states nonlocally in distant emitter-cavity systems, such as atom-cavity system \cite{atomghz} and nitrogen vacancy (N-V) centers-photonic crystal (PC) nano-cavity system \cite{nvghz}.
The GHZ state is one of the most important multiparticle quantum resources because its potential applications in quantum
information processing such as quantum teleportation \cite{ghztele}, entanglement swapping
\cite{ghzswap}, fundamental test of quantum mechanics \cite{ghztest} and quantum cryptographic\cite{densecoding,xiayanhgsa}. In most of these applications, the efficient measurement and analysis of entangled states are essential and demanding.

The complete and deterministic analysis of the two-qubit and multi-qubit entangled states is required as a vital element for many important applications in quantum communication, such as quantum teleportation \cite{teleporation,teleporation1,teleporation2,ghztele}, quantum dense
coding \cite{densecoding,wangcoding}, quantum superdense coding \cite{superdensecoding}, and so on. Many approaches have been investigated
for two-qubit and three-qubit entangled analyzers. For example, only with the help of linear optics, the probability for analysis of Bell-states is 50\% and the probability for analyzing GHZ states
(GSA) is 25\% \cite{GHZA}. However, Hyperentanglement allows the complete analysis of entanglement state becomes possible. Hyperentanglement is the entanglement of photons simultaneously existing in more than one degree of freedom (DOF). In
2003, Walborn \emph{et al.}  \cite{HyperentangledBell-stateanalysis2003} proposed a simple linear-optical scheme for the complete Bell-state
measurement of photons by using hyperentanglement. In 2006,
 Carsten Schuck \emph{et al.} \cite{HyperentangledBell-stateanalysis2006}
deterministically distinguished all four polarization Bell
states of entangled photon pairs with the aid of polarization-time-bin
hyperentanglement in an experiment. In 2007, Barbieri \emph{et al.}
\cite{HyperentangledBell-stateanalysis2007p-m} realized
complete and deterministic Bell-state measurement in the experiment using
linear optics and two-photon polarization-momentum hyperentanglement.
In 2012, Song \emph{et al.} \cite{song} proposed a scheme to
distinguish eight GHZ states completely using polarization-spatial-mode hyperentanglement with
only linear optics.

Aside from the complete analysis of entangled states, other applications of hyperentanglement have been extensively studied because it can improve the channel
capacity of long-distance quantum communications and offers significant advantages in quantum communication protocols. For example, hyperentanglement can be used for quantum error-correcting
code \cite{hyperentanglement-app-error-correctingcode}, high-capacity quantum cryptography \cite{wangcoding, hyperentanglement-app-c}, quantum repeaters \cite{wangtjpra}, and deterministic entanglement purification \cite{hpur}.  The 16 hyperentangled two-qubit Bell states \cite{shengprainpress} and 64 three-qubit GHZ states \cite{xiayanhgsa} may be completely distinguished with the
help of cross-Kerr nonlinearity in the optical single-photon regime. However, due to the weak cross-Kerr nonlinearity, the schemes are currently without technical support \cite{44,45}.

\begin{figure}[!ht]
\begin{center}
\begin{center}
\includegraphics[width=10cm,angle=0]{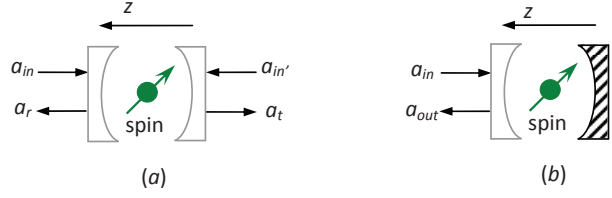}
\caption{ A charged quantum dot inside a double-sided (a) or in a single-sided (b) micropillar cavity.
An exciton consisting of two electrons bound to one hole with negative charges can be created
by the optical excitation of a photon and an electron-spin.} \label{fig1}
\end{center}
\end{center}
\end{figure}

Embedding quantum dots \cite{QD} (QDs) into solid-state cavities is relatively easy, and the deterministic transfer
of quantum information between photons and spins in QDs can be promoted by two typical structures of cavity-QD systems, \emph{i.e.}, the double-sided cavity-QD system and the single-sided cavity-QD system (see in Fig.\ref{fig1}). These two spin-cavity systems can be used to produce entangled photon-states, such as the Bell state, the GHZ state, and cluster state\cite{7,hybrid,9} and perform Bell-state \cite{6,13} and hyperentangled Bell-state \cite{renoehbsa,Wangprahbsa} analysis. Double-sided-cavity-spin systems have been proven to work like a beam splitter when limited by a weak incoming field \cite{beamsplitter} in a weak-coupling cavity where the vacuum Rabi frequency is less than the
cavity decay rate. Thus, the double-sided cavity-spin unit can be used as a SWAP gate between the polarization DOF and the spatial-mode DOF of photons and records the relationship between the phase information in these DOFs \cite{Wangprahbsa}. The single-sided cavity-QD unit can be used as a parity checking device that will pick up a phase $\theta$ on the photon if this photon is coupled with the spin in QD, and the $\theta$ is turnable \cite{13}.
Because of its high quantum efficiency, single-photon characteristics, and high stability, the spin-photon interface has been widely studied in deterministic optical quantum computing \cite{10}, construction universal gates \cite{6,7,9}, hybrid entanglement generation \cite{hybrid}, and quantum purification and concentration \cite{WANGCHUAN,wangoe}.

In this paper, we take the advantages of these two spin-cavity units and propose a scheme that can be used as a complete three-qubit hyperentangled-GHZ-state analyzer (HGSA). The scheme can be used to generate 3-photon hyperentangled GHZ states and increase the channel capacity of long-distance quantum
communication. The proposed HGSA device can be applied in the crucial components of long-distance quantum communication, such as high-capacity teleportation, entanglement swapping, and some important quantum cryptographic schemes \cite{xiayanhgsa}. As the GHZ state is the extension of two-qubit Bell state to multi-qubit system, this device can also be used to analysis hyperentangled Bell-state photon-pairs. Compared with the previous work \cite{renoehbsa,Wangprahbsa}, this scheme has some advantages.
First, it is much simpler than those schemes because only two spin-cavity units are employed here, not four or more.
Second, it can be generalized to N-photon hyperentangled GHZ states analysis, and the number of the required spin-cavity units will not increase with the number of the photons. Moreover, it can be used to generate hyperentangled GHZ states.
 With existing experimental data, we demonstrate that the present scheme can work in the strong-coupling regime using current technologies.

\section{ The complete analysis of hyperentangled three-photon GHZ states in two DOFs using quantum-dot spin and optical double-sided microcavity system}

\begin{figure}[!ht]
\begin{center}
\includegraphics[width=8.5cm,angle=0]{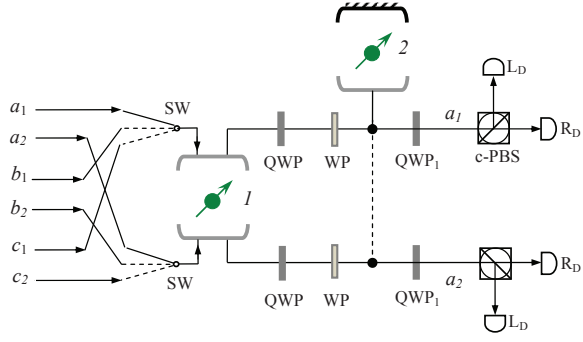}
\caption{ The setup for determining the phase information in polarization and spatial-mode DOFs. QWP represents a quarter-wave plate
that can be used to accomplish Hadamard operations on the $\vert R\rangle$ and
$\vert L\rangle$ polarized states: $\vert R\rangle\rightarrow\frac{1}{\sqrt{2}}(\vert R\rangle+\vert L\rangle)$ and
$\vert L\rangle\rightarrow\frac{1}{\sqrt{2}}(\vert R\rangle-\vert L\rangle)$. QWP$_1$ represents the quarter-wave plate which is different from QWP
and can be used to accomplish another Hadamard operation: $\vert R\rangle\rightarrow\frac{1}{\sqrt{2}}(\vert R\rangle-\vert L\rangle)$ and
$\vert L\rangle\rightarrow\frac{1}{\sqrt{2}}(\vert R\rangle+\vert L\rangle)$. WP is used to accomplish the single photon operation $\sigma^i$ on polarized states: $\sigma^i\vert R\rangle\rightarrow\vert R\rangle$ and $\sigma^i\vert
L\rangle\rightarrow i\vert L\rangle$. c-PBS is a polarizing beam splitter in the circular
basis which transmits the input right-circularly
polarized photon $|R\rangle$ and reflects the left-circularly
polarized photon $|L\rangle$.} \label{fig.2}
\end{center}
\end{figure}

In this section, we introduce the principle of the complete analysis of hyperentangled three-photon GHZ state in polarization and the spatial-mode DOFs using quantum-dot spin and optical double-sided microcavity system. A quantum system with three photons hyperentangled
in polarization and spatial-mode DOFs has 64 generalized GHZ states that can be expressed as
\begin{eqnarray}
|\varphi^{ABC}\rangle_{PS}=|\Psi_{ABC}\rangle_P\otimes|\Phi_{ABC}\rangle_S
\end{eqnarray}
where subscripts $A$, $B$ and $C$ represent the three photons
in the hyperentangled state. The subscript $P$ denotes the polarization DOF of the photons and $|\Psi_{ABC}\rangle_P$ is one of the eight Bell states in the
polarization DOF, which is expressed as
\begin{eqnarray}
|\Psi^{\pm}_{1}\rangle_{P}=\frac{1}{\sqrt{2}}(|RRR\rangle_{ABC}\pm|LLL\rangle_{ABC}),\;\;
|\Psi^{\pm}_{2}\rangle_{P}=\frac{1}{\sqrt{2}}(|RRL\rangle_{ABC}\pm|LLR\rangle_{ABC}),\nonumber\\
|\Psi^{\pm}_{3}\rangle_{P}=\frac{1}{\sqrt{2}}(|RLR\rangle_{ABC}\pm|LRL\rangle_{ABC}),\;\;
|\Psi^{\pm}_{4}\rangle_{P}=\frac{1}{\sqrt{2}}(|LRR\rangle_{ABC}\pm|RLL\rangle_{ABC}).
\end{eqnarray}
Here, $\vert L \rangle $ and $\vert R
\rangle$ represent the left and right
circular polarizations of photons, respectively.
The subscript $S$ denotes the spatial-mode DOF and $|\Phi_{ABC}\rangle_S $ is one of the eight Bell states in the spatial-mode DOF, which is expressed as
\begin{eqnarray}
|\Phi^{\pm}_{1}\rangle_{S}=\frac{1}{\sqrt{2}}(|a_1b_1c_1\rangle_{ABC}\pm|a_2b_2c_2\rangle_{ABC}),\;\;
|\Phi^{\pm}_{2}\rangle_{S}=\frac{1}{\sqrt{2}}(|a_1b_1c_2\rangle_{ABC}\pm|a_2b_2c_1\rangle_{ABC}),\nonumber\\
|\Phi^{\pm}_{3}\rangle_{S}=\frac{1}{\sqrt{2}}(|a_1b_2c_1\rangle_{ABC}\pm|a_2b_1c_2\rangle_{ABC}),\;\;
|\Phi^{\pm}_{4}\rangle_{S}=\frac{1}{\sqrt{2}}(|a_2b_1c_1\rangle_{ABC}\pm|a_1b_2c_2\rangle_{ABC}),
\end{eqnarray}
where $i_1 $ and $i_2$ ($i$ is $a,$ $b,$ or $c$) are the different spatial modes for
the photon $I$ ($I$ is corresponding photon $A$, $B,$ or $C$).
Consider that the three hyperentangled photons $ABC$ is in one of the 64 hyperentangled GHZ states that forms in Eq. (1).
The principles of the complete three-photon HGSA is shown in Fig. \ref{fig.2} .

\subsection{Measurement of the relation of the phase information between two DOFs by using quantum-dot spin and optical double-sided microcavity system}

The first cavity (cavity 1) is a double-sided optical microcavity in which the top and bottom mirrors are both partially reflective, as shown in Fig. 1 (a), and a singly charged QD is embedded in cavity 1. The spin of this QD (spin 1) is initialized in the state $|\Psi^s\rangle_1=(|\uparrow\rangle+|\downarrow\rangle)_1/\sqrt{2}$. Pauli's exclusion principle illustrates that, the $|\uparrow\rangle$ spin state ($J_z=+\frac{1}{2}$) only couples with the $s_z=1$ photon, \emph{ i.e.}, the $|R^{\uparrow}\rangle$ or $|L^{\downarrow}\rangle$ photon, to the exciton X$^-$ in state $|\uparrow\downarrow\Uparrow\rangle$. When the electron spin is in the spin state $|\downarrow\rangle$ $(J_z=-\frac{1}{2})$, it only couples the photon that is in the state $|R^{\downarrow}\rangle$ or $|L^{\uparrow}\rangle (s_z=-1)$ to the exciton X$^-$ in state $|\downarrow\uparrow\Downarrow\rangle$. Here, $|\Uparrow\rangle$ and $|\Downarrow\rangle$ represent heavy-hole spin states with $J_z=\frac{3}{2}$ and $J_z=-\frac{3}{2}$, respectively, and the superscript arrows $\uparrow$ and $ \downarrow$ in the photon state indicate the propagation
direction along the $z$ axis.
For a double-sided cavity, the coupling photon will be reflected by the cavity and the uncoupling photon will pass through the cavity.
All reflection and transmission coefficients of this spin-cavity system can be
determined by solving the Heisenberg equations of motion for
the cavity-field operator ($\hat{a}$) and the exciton $X^-$ dipole operator ($\sigma_{-}$) in weak
excitation approximation \cite{6,7,hybrid,9}
\begin{eqnarray}
\frac{d \hat{a}}{dt}&=&-[i (\omega_c-\omega)+\kappa+\frac{\kappa_s}{2}]\hat{a}-g \sigma_{-}-\sqrt{\kappa}\hat{a}_{in'}-\sqrt{\kappa}\hat{a}_{in}+\hat{H},\nonumber\\
\frac{d \sigma_{-}}{dt}&=&-[i (\omega_{X^-}-\omega)+\frac{\gamma}{2}]\sigma_{-}-g \sigma_{z}\hat{a}+\hat{G},\nonumber\\
\;\;\hat{a}_r\;&=&\hat{a}_{in}+\sqrt{\kappa}\hat{a},\;\;\;\hat{a}_t\;=\hat{a}_{in'}+\sqrt{\kappa}\hat{a},
\end{eqnarray}
where $\omega$, $\omega_c$, and $\omega_{X^-}$ are the frequencies of the photon, the cavity and $X^-$ transition, respectively; $g$ represents the coupling constant; $\gamma$
is the exciton dipole decay rate; and $\kappa$ and $\kappa_s$ are
the cavity decay rate and the leaky rate, respectively. $\hat{H}$ and $\hat{G}$ are
the noise operators related to reservoirs. $\hat{a}_{in}$, $\hat{a}_{in'}$ and $\hat{a}_r$, $\hat{a}_t$ are
the input and output field operators. For a double-sided optical microcavity system \cite{beamsplitter,hybrid}, the reflection $r_h$ and transmission $t_h$ coefficient in a coupled cavity in the resonant interaction case can be described by
\begin{eqnarray}
r_h(\omega)&=&1+t_h(\omega),\nonumber\\
t_h(\omega)&=&-\frac{\kappa[i(\omega_{X^{-}}-\omega)+\frac{\gamma}{2}]}{[i(\omega_{X^{-}}-\omega)
+\frac{\gamma}{2}][i(\omega_c-\omega)+\kappa+\frac{\kappa_s}{2}]+g^2}.
\end{eqnarray}
When $g=0$, the reflection $r_0$ and transmission $t_0$ coefficients in an uncoupled cavity are
\begin{eqnarray}
r_0(\omega)=\frac{i(\omega_0-\omega)+\frac{\kappa_s}{2}}{i(\omega_c-\omega)+\kappa+\frac{\kappa_s}{2}},\;\;
t_0(\omega)=\frac{-\kappa}{i(\omega_0-\omega)+\kappa+\frac{\kappa_s}{2}}.
\end{eqnarray}

Thus, when the side leakage and cavity loss ($k_s$) can be ignored, in the resonant interaction case, the cold
(uncoupled) and hot (coupled) cavities generally have different
reflection and transmission coefficients, and the dynamics of the photon and of the spin in the cavity are described as follows:
\begin{eqnarray}
& &\vert R^{\uparrow}, \uparrow \rangle \rightarrow \vert
L^{\downarrow}, \uparrow\rangle, \;\;\;\; \vert
L^{\uparrow}, \uparrow
\rangle \rightarrow -\vert L^{\uparrow}, \uparrow\rangle,\;\;\;\;
\vert R^{\downarrow}, \uparrow \rangle\rightarrow -\vert
R^{\downarrow},\uparrow \rangle,  \;\;\;\;  \vert
L^{\downarrow},\uparrow
\rangle\rightarrow \vert R^{\uparrow},\uparrow \rangle, \nonumber\\
& &\vert R^{\uparrow}, \downarrow \rangle\rightarrow -\vert
R^{\uparrow},\downarrow \rangle, \;\;\;\;  \vert
L^{\uparrow}, \downarrow \rangle\rightarrow \vert
R^{\downarrow},\downarrow \rangle,\;\;\;\;
\vert R^{\downarrow}, \downarrow \rangle\rightarrow \vert
L^{\uparrow},\downarrow \rangle,  \;\;\;\;   \vert
L^{\downarrow}, \downarrow \rangle\rightarrow -\vert
L^{\downarrow},\downarrow \rangle.\label{5}
\end{eqnarray}
From the Eq. (\ref{5}), one can see that for the coupling case, the photon will be reflected by the cavity and the polarized state of the photon will be flipped for the unchanged photon spin $s_z$. For the uncoupling case, the photon will only add a $\pi$ phase when it passes through the cavity. The photon polarization and electron spin may become entangled when this
spin-cavity unit is used.

Using the spin-cavity unit discussed above, one can record the relation between two DOFs. After the interaction between the spin 1 and photons $ABC$, the state in this two DOFs can be exchanged, and the phase information in this two DOfs will be flipped.
First, let photons $A$, $B$, and $C$ successively pass through the cavity 1 from the left input-port. A time interval $\triangle t$ exists between photons $A$ and $B$ and between photons $B$ and $C$. $2\triangle t$ should be less than the spin coherence
time $T$.

According to the evolution rules of the photon and the spin in the cavity described in Eq. (\ref{5}), after three photons $ABC$ passes through cavity 1 and the QWPs, the whole system of the three photons and spin 1 evolves as
\begin{eqnarray}
|\Psi^{\pm}_{i}\rangle_{P}|\Phi^{\pm}_{j}\rangle_{S}|+\rangle^s_1& \rightarrow&|\Psi^{\mp}_{j}\rangle_{P}|\Phi^{\mp}_{i}\rangle_{S}|-\rangle^s_1,\nonumber\\
|\Psi^{\mp}_{i}\rangle_{P}|\Phi^{\pm}_{j}\rangle_{S}|+\rangle^s_1&\rightarrow&|\Psi^{\mp}_{j}\rangle_{P}|\Phi^{\pm}_{i}\rangle_{S}|+\rangle^s_1,
\end{eqnarray}
where $|\pm\rangle=\frac{1}{\sqrt{2}}(|\uparrow\rangle\pm|\downarrow\rangle)$, and the overall phase of the system is ignored.

Equation (\ref{8}) shows that when three photons pass through a double-sided cavity, the states in the polarization DOF and the states in the spatial-mode DOF of the photons $A$, $B$, and $C$ will be interchanged and the phase information in these different DOFs will be flipped. That is, the double-sided cavity works as a SWAP-like gate for these two DOFs. At the same time, the relation between the phase information in these two DOFs will be stored in spin 1. If the phase information in the two DOFs are the same (\emph{i.e.}, the hyperentangled state is in $\vert \Psi^{\pm}_{i}\rangle_P\vert\Phi^{\pm}_{j} \rangle_S, i, j=1, 2, 3$, and $4$), the state of the spin 1 changes to $|-\rangle$ when the photons ABC pass through the double-sided cavity.
Otherwise, the spin 1 remains in the state $|+\rangle$ when the phase information in the two DOFs differs (the hyperentangled state is in $\vert \Psi^{\pm}_{i}\rangle_P\vert\Phi^{\mp}_{j} \rangle_S$). Therefore, the spin states divide the states
of the $ABC$ photons into two groups: $\{ |\Psi^{\pm}_{ABC}\rangle_P|\Phi^{\pm}_{ABC}\rangle_S,|\Psi^{\pm}_{ABC}\rangle_P|\Phi^{\mp}_{ABC}\rangle_S\}$.
Note that, the states in the two different DOFs are interchanged and the phase information in these two DOFs is flipped after the interaction between the photons and spin 1 in the first double-sided cavity.

\subsection{Determination of the phase information in polarization DOF based on quantum-dot spin and optical single-sided microcavity system}

The second cavity which also involves a singly charged QD embedded in is a single-sided microcavity with the top mirror partially reflective and the bottom mirror 100$\%$ reflective, as shown in Fig.2. The spin of the QD (spin 2) is also prepared in the state $|\Psi_2\rangle=(|\uparrow\rangle+|\downarrow\rangle)_1/\sqrt{2}$. The different reflection coefficients for the coupling ($r_h'$) and uncoupling spin-cavity systems ($r_0'$) can be determined by solving another Heisenberg equations of motion \cite{6,7,13}, and
\small
\begin{eqnarray}
r_h'(\omega)&=&1-\frac{\kappa[i(\omega_{X^{-}}-\omega)+\frac{\gamma}{2}]}{[i(\omega_{X^{-}}-\omega)
+\frac{\gamma}{2}][i(\omega_c-\omega)+\frac{\kappa}{2}+\frac{\kappa_s}{2}]+g^2},\nonumber\\
r_0'(\omega)&=&\frac{i(\omega_0-\omega)+\frac{\kappa_s}{2}}{i(\omega_c-\omega)+\frac{\kappa}{2}+\frac{\kappa_s}{2}}.
\end{eqnarray}\normalsize
By suitable detuning of the photon frequency, the phase difference between $r_h'$ and $r_0'$
can be adjusted to $\pi/2$, and a photon-spin phase-shift gate can be developed\cite{7}as follows:
\begin{eqnarray}
\vert R, \uparrow \rangle \rightarrow \vert
L, \uparrow\rangle,\;\; \vert L, \uparrow
\rangle \rightarrow i \vert R, \uparrow\rangle,\nonumber\\
\vert R, \uparrow \rangle\rightarrow i\vert
L,\uparrow \rangle, \;\;\vert L,\uparrow
\rangle\rightarrow \vert R,\uparrow \rangle.\label{51}
\end{eqnarray}
From the Eq. (\ref{51}), one can see that in the coupling case, the photon will be reflected by the cavity and the state of the photon will be flipped for the unchanged photon spin $s_z$. In the uncoupling case, the photon will not only be flipped but also pick up a $\pi/2$ phase when it is reflected by the cavity.
 This type of spin-cavity unit can be used to determine the phase information in the polarization DOF without destroying the photons $ABC$\cite{renoehbsa,13}.

As shown in Fig. \ref{fig.2}, after passing through the QWPs, which act as
the Hadamard operations on the polarization DOF, that is,
\begin{eqnarray}
\vert R \rangle  \rightarrow (\vert  R \rangle+\vert  L \rangle)/\sqrt{2},\;\;\;\;
\vert L \rangle  \rightarrow (\vert  R \rangle-\vert  L \rangle)/\sqrt{2},
\end{eqnarray}
the eight GHZ states $|\Psi^{\pm}_{j}\rangle_P$ ($j=1, 2, 3, 4$) in the
polarization DOF evolve as
\begin{eqnarray}
|\Psi^{+}_{1}\rangle_{P}&\rightarrow&\frac{1}{2}(|RRR\rangle+|RLL\rangle+|LRL\rangle+|LLR\rangle)_{ABC}=|\psi^{+}_{1}\rangle_{P},\nonumber\\
|\Psi^{-}_{1}\rangle_{P}&\rightarrow&\frac{1}{2}(|LLL\rangle+|LRR\rangle+|RLR\rangle+|RRL\rangle)_{ABC}=|\psi^{-}_{1}\rangle_{P},\nonumber\\
|\Psi^{+}_{2}\rangle_{P}&\rightarrow&\frac{1}{2}(|RRR\rangle+|RLL\rangle-|LRL\rangle-|LLR\rangle)_{ABC}=|\psi^{+}_{2}\rangle_{P},\nonumber\\
|\Psi^{-}_{2}\rangle_{P}&\rightarrow&-\frac{1}{2}(|LLL\rangle+|LRR\rangle-|RLR\rangle-|RRL\rangle)_{ABC}=|\psi^{-}_{2}\rangle_{P},\nonumber\end{eqnarray}
\begin{eqnarray}
|\Psi^{+}_{3}\rangle_{P}&\rightarrow&\frac{1}{2}(|RRR\rangle-|RLL\rangle+|LRL\rangle-|LLR\rangle)_{ABC}=|\psi^{+}_{3}\rangle_{P},\nonumber\\
|\Psi^{-}_{3}\rangle_{P}&\rightarrow&-\frac{1}{2}(|LLL\rangle-|LRR\rangle+|RLR\rangle-|RRL\rangle)_{ABC}=|\psi^{-}_{3}\rangle_{P},\nonumber\\
|\Psi^{+}_{4}\rangle_{P}&\rightarrow&\frac{1}{2}(|RRR\rangle-|RLL\rangle-|LRL\rangle+|LLR\rangle)_{ABC}=|\psi^{+}_{4}\rangle_{P},\nonumber\\
|\Psi^{-}_{4}\rangle_{P}&\rightarrow&\frac{1}{2}(|LLL\rangle-|LRR\rangle-|RLR\rangle+|RRL\rangle)_{ABC}=|\psi^{-}_{4}\rangle_{P}.\label{Hadamard}
\end{eqnarray}

From Eq. (\ref{Hadamard}), we can see that, after the Hadamard transformation, the number of $|R\rangle$ is odd for the states $|\Psi^{+}_{j}\rangle_P$ with the superscript $+$ while the number of $|R\rangle$ is even for other states,\emph{ i.e.}, $|\Psi^{-}_{j}\rangle_P$.
The eight GHZ states $|\Psi^{\pm}_{j}\rangle_P$ can be divided into two groups: $\{ |\Psi^{+}_{ABC}\rangle_P,|\Psi^{-}_{ABC}\rangle_P$. With grouping and the single-sided cavity-QD unit which is same as that by Ren et al.\cite{renoehbsa}, we can complete the task of phase-determination in the polarization DOF.
From Fig. 2, when three photons are reflected by cavity 2 and gain a $\frac{\pi}{2}$ phase on all of the $|L\rangle$ photons using wave planes (WPs), the total states of the three photons with one spin
are transformed into
\begin{eqnarray}
|\psi^{+}_{1}\rangle_{P}|+\rangle^s_2&\rightarrow&\frac{-i}{2}(|LLL\rangle+|LRR\rangle+|RLR\rangle+|RRL\rangle)_{ABC}|+'\rangle^s_2,\nonumber\\
|\psi^{-}_{1}\rangle_{P}|+\rangle^s_2&\rightarrow&\frac{-i}{2}(|RRR\rangle+|RLL\rangle+|LRL\rangle+|LLR\rangle)_{ABC}|-'\rangle^s_2,\nonumber\\
|\psi^{+}_{2}\rangle_{P}|+\rangle^s_2&\rightarrow&\frac{-i}{2}(|LLL\rangle+|LRR\rangle-|RLR\rangle-|RRL\rangle)_{ABC}|+'\rangle^s_2,\nonumber\\
|\psi^{-}_{2}\rangle_{P}|+\rangle^s_2&\rightarrow&\frac{i}{2}(|RRR\rangle+|RLL\rangle-|LRL\rangle-|LLR\rangle)_{ABC}|-'\rangle^s_2,\nonumber\\
|\psi^{+}_{3}\rangle_{P}|+\rangle^s_2&\rightarrow&\frac{-i}{2}(|LLL\rangle-|LRR\rangle+|RLR\rangle-|RRL\rangle)_{ABC}|+'\rangle^s_2,\nonumber\\
|\psi^{-}_{3}\rangle_{P}|+\rangle^s_2&\rightarrow&\frac{i}{2}(|RRR\rangle-|RLL\rangle+|LRL\rangle-|LLR\rangle)_{ABC}|-'\rangle^s_2,\nonumber\\
|\psi^{+}_{4}\rangle_{P}|+\rangle^s_2&\rightarrow&\frac{-i}{2}(|LLL\rangle-|LRR\rangle-|RLR\rangle+|RRL\rangle)_{ABC}|+'\rangle^s_2,\nonumber\\
|\psi^{-}_{4}\rangle_{P}|+\rangle^s_2&\rightarrow&\frac{-i}{2}(|RRR\rangle-|RLL\rangle-|LRL\rangle+|LLR\rangle)_{ABC}|-'\rangle^s_2.
\end{eqnarray}
Here, $|+'\rangle^s_2=\frac{1}{\sqrt{2}}(|\uparrow\rangle+i|\downarrow\rangle)$ and $|-'\rangle^s_2=\frac{1}{\sqrt{2}}(|\uparrow\rangle-i|\downarrow\rangle)$.
With an measurement on the spin 2 in basis $\{|+'\rangle, |-'\rangle\}$, one can
read out the information about the phases in the polarization DOF. If the spin 2 is in the state $|+'\rangle^s_2$, the phase information is $+$ in polarization DOF; otherwise, the phase information is $-$, when the measuring result of the spin 2 is $|-'\rangle^s_2$. The relation between the measurement outcomes of spins $1$ and $2$ and the
initial phase information of the hyperentangled photons system is shown in Table \ref{tab1}.

From Table \ref{tab1}, one can see that the phase information in two DOFs can be obtained by measuring spins 1 and 2. Electron spin $1$ is measured using the spin basis $\{|+\rangle, |-\rangle\}$ whereas electron spin $2$ is measured using the spin basis $\{|+'\rangle, |-'\rangle\}$. Spin $1$ records the relation between the phase information in the polarization and spatial-mode DOFs, while spin 2 can be used to determine the initial information encoded in the spatial-mode DOF. In detail, when spin 2 is detected in the state $|+'\rangle$, the initial phase information of the photons in the spatial-mode DOF is $-$. When the spin 2 is detected in the state $|-'\rangle$, the photons are initially in the spatial-mode state $|\Phi^{+}\rangle_S$. The phase information in the polarization DOF can be obtained when the phase information in the spatial-mode DOF is determined and spin 1 is measured using $\{|+\rangle,|-\rangle\}$ as the basis. If spin 1 is in the state $|-\rangle$, the phase information in two DOFs is same; otherwise, the phase information in two DOFs is different when the spin 1 is in the state $|+\rangle$.

\begin{table}[!ht]
\begin{center}
 \caption{Relationship between the measurement outcomes of two spins and the initial phase information of the hyperentangled states}
\begin{tabular}{ccccccccccc}\hline\hline
& Spin $1$ and $2$ && the initial phase information & \\\hline & $|+\rangle_1|+'\rangle_2$ & &
$|\Psi^{-}\rangle_P\otimes|\Phi^{+}\rangle_S$ \\
& $|+\rangle_1|-'\rangle_2$& &
$|\Psi^{+}\rangle_P\otimes|\Phi^{-}\rangle_S$\\
&
$|-\rangle_1|+'\rangle_2$ & &
$|\Psi^{-}\rangle_P\otimes|\Phi^{-}\rangle_S$ \\
&
$|-\rangle_1|-'\rangle_2$ & &
$|\Psi^{+}\rangle_P\otimes|\Phi^{+}\rangle_S$ \\
\hline\hline
\end{tabular}\label{tab1}
\end{center}
\end{table}

Then, by applying Hadamard transformations
on the three photons in the polarization DOF, the the polarization states of the photons $ABC$ are changed back into the standard GHZ-state form. Finally, the photons $ABC$ can be independently measured in both the
polarization and spatial-mode DOFs with
single-photon detectors.
Thus, as shown in Table \ref{tab2} and \ref{tab3}, the outcomes of three photon polarization states determine the parity information in the spatial-mode DOF, whereas the outcomes of the spatial-mode states determine the parity information in the
polarization DOF. The relationship between the measurement outcomes of the states of the three photons $ABC$ and the initial parity information of hyperentangled photon system is shown in Table \ref{tab2} and \ref{tab3}.

\begin{table}[!ht]
\begin{center}
\caption{Relationship between the final polarization states of three photons and the initial hyperentangled states.}
\begin{tabular}{ccccccccccc}\hline\hline
& the final Spatial-mode state & & the initial Polarization state &  \\\hline
& $\vert
RRR\rangle_{ABC}$, $\vert LLL\rangle_{ABC}$ & &$|\Phi^{\pm}_{1}\rangle_P$ \\
& $\vert
RRL\rangle_{ABC}$, $\vert LLR\rangle_{ABC}$& &
$|\Phi^{\pm}_{2}\rangle_P$ \\
& $\vert
RLR\rangle_{ABC}$, $\vert LRL\rangle_{ABC}$& &
$|\Phi^{\pm}_{3}\rangle_P$ \\
& $\vert
LRR\rangle_{ABC}$, $\vert RLL\rangle_{ABC}$ & & $ |\Phi^{\pm}_{4}\rangle_P$ \\
\hline\hline
\end{tabular}\label{tab2}
\end{center}
\end{table}

\begin{table}[!ht]
\begin{center}
 \caption{Relationship between the final Spatial-mode states of three photons and the initial hyperentangled states.}
\begin{tabular}{ccccccccccc}\hline\hline
& the final Spatial-mode state & & the initial Polarization state &  \\\hline
& $\vert
a_1 a_1 a_1\rangle_{ABC}$, $\vert a_2 a_2a_2\rangle_{ABC}$ & &$|\Psi^{\pm}_{1}\rangle_P$ \\
& $\vert
a_1 a_1 a_2\rangle_{ABC}$, $\vert a_2 a_2 a_1\rangle_{ABC}$& &
$|\Psi^{\pm}_{2}\rangle_P$ \\
& $\vert
a_1 a_2 a_1\rangle_{ABC}$, $\vert a_2 a_1 a_2\rangle_{ABC}$& &
$|\Psi^{\pm}_{3}\rangle_P$ \\
& $\vert
a_2 a_1 a_1\rangle_{ABC}$, $\vert a_1 a_2 a_2\rangle_{ABC}$ & & $ |\Psi^{\pm}_{4}\rangle_P$ \\
\hline\hline
\end{tabular}\label{tab3}
\end{center}
\end{table}

According to the measurement of the final states of photons $ABC$ in the
polarization and spatial-mode DOFs and the detection result of the spins $1$ and $2$, the initial hyperentangled-GHZ state of the three
photons $ABC$ can be determined. From the analysis above, one can see that three-photon
64 GHZ states can be completely distinguished with
the help of QD-cavity systems. Moreover, this device can be generalized to N-photon hyperentangled GHZ states analysis and it can also be used to analysis hyperentangled Bell-state photon-pairs. Compared with the previous works\cite{renoehbsa}, our scheme is much simpler as only two spin-cavity units are used here, not four or more. The two QD spins accomplish the task of phase measurement. With the help of the linear wave plates (QWPs and WPs) and two
different spin-cavity units, the phase information in two different DOFs can be obtained without destroying
three photons. It shows that our device in Fig. \ref{fig.2} can theoretically accomplish a complete and deterministic HGSA with 100$\%$ success probability.

\section{Application of the HGSA device in quantum communication}

In the previous section, we used two different spin-cavity units to construct a
device completing a HGSA. Now, we introduce the application of this device in high-capacity, long-distance quantum communication.

\begin{figure}[!ht]
\begin{center}
\includegraphics[width=8.5cm,angle=0]{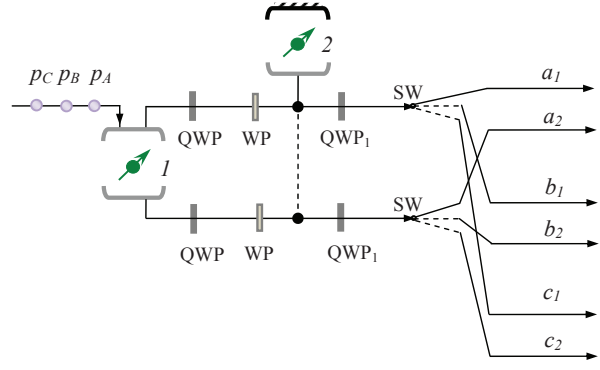}
\caption{Schematic diagram of hyperentanglement GHZ state generation.}
\label{fig.3}
\end{center}
\end{figure}
One application of this HGSA device is the generation of hyperentangled-GHZ-state photons.
The principle of the hyperentangled GHZ state photon generation (HGSG) is shown in Fig. \ref{fig.3}. Three photons, $A$, $B$, and $C$, are prepared in the same initial state
$| \varphi_{A}\rangle=| \varphi_{B}\rangle=| \varphi_{C}\rangle=|R\rangle$, and two spins ($1$ and $2$) are prepared in the state $|+\rangle_{1}=|+\rangle_{2}=\frac{1}{\sqrt{2}}(|\uparrow\rangle+|\downarrow\rangle)$. Spin 1 is embedded in a double-sided optical microcavity 1, and spin 2 is embedded in a single-sided optical microcavity 2.
Fig. \ref{fig.3} shows that photons $A$, $B$ and $C$ are successively sent into the cavities from the left input port of the HBSG device. After passing through two cavities, the photons can become entangled with the electron-spins. The evolution of the entire system can then be described as
\begin{eqnarray}
&&|\varphi_{p_A}\rangle|\varphi_{p_B}\rangle|\varphi_{p_C}\rangle|+\rangle_{1}|+\rangle_{2}\nonumber\\
&^{\underrightarrow{cavity1}}&\frac{1}{2}|+\rangle_2[|-\rangle_1(\vert \Psi^-_{1}\rangle_P\vert\Phi^-_{1} \rangle_S-\vert \Psi^+_{1}\rangle_P\vert\Phi^+_{1} \rangle_S)_{ABC}\nonumber\\&&+\;\;|+\rangle_1(\vert \Psi^+_{1}\rangle_P\vert\Phi^-_{1} \rangle_S-\vert \Psi^-_{1}\rangle_P\vert\Phi^+_{1} \rangle_S)_{ABC}]\nonumber\\
&^{\underrightarrow{QWP}}&\frac{1}{2}|+\rangle_2[|-\rangle_1(\vert \psi^-_{1}\rangle_P\vert\Phi^-_{1} \rangle_S-\vert \psi^+_{1}\rangle_P\vert\Phi^+_{1} \rangle_S)_{ABC}\nonumber\\&&+\;\;|+\rangle_1(\vert \psi^+_{1}\rangle_P\vert\Phi^-_{1} \rangle_S-\vert \psi^-_{1}\rangle_P\vert\Phi^+_{1} \rangle_S)_{ABC}]\nonumber\\
&^{\underrightarrow{WP,\;cavity2}}&\frac{-i}{2}(|-\rangle_1|-'\rangle_2\vert \psi^+_{1}\rangle_{ABC_P}\vert\Phi^-_{1} \rangle_{ABC_S}-|-\rangle_1|+'\rangle_2\vert \psi^-_{1}\rangle_{ABC_P}\vert\Phi^+_{1}\rangle_{ABC_S}\nonumber\\&&+\;\;|+\rangle_1|+'\rangle_2\vert\vert \psi^-_{1}\rangle_{ABC_P}\vert\Phi^-_{1}\rangle_{ABC_S}-|+\rangle_1|-'\rangle_2\vert\vert \psi^+_{1}\rangle_{ABC_P}\vert\Phi^+_{1} \rangle_{ABC_S})\nonumber\\
&^{\underrightarrow{QWP_1}}&\frac{-i}{2}(|-\rangle_1|-'\rangle_2\vert \Psi^-_{1}\rangle_{ABC_P}\vert\Phi^-_{1}\rangle_{ABC_S}-|-\rangle_1|+'\rangle_2\vert \Psi^+_{1}\rangle_{ABC_P}\vert\Phi^+_{1}\rangle_{ABC_S}\nonumber\\&&+\;\;|+\rangle_1|+'\rangle_2\vert\vert \Psi^+_{1}\rangle_{ABC_P}\vert\Phi^-_{1}\rangle_{ABC_S}-|+\rangle_1|-'\rangle_2\vert\vert \Psi^-_{1}\rangle_{ABC_P}\vert\Phi^+_{1} \rangle_{ABC_S}), \label{8}
\end{eqnarray}

Eq. (\ref{8}) shows that the spins $1$ and $2$ in the cavities record the phase information in these two DOFs. If spin 2 is in the state $|+'\rangle$, the phase information in spatial-mode DOF is `+'; otherwise the phase information in the spatial-mode DOF is `-' when spin 2 is in the state $|-'\rangle$. If the final state of the photon is $\vert \Psi^{\pm}_{1}\rangle_P\vert\Phi^{\pm}_{1} \rangle_S $,\emph{ i.e.}, the phase information in two DOFs is the same, the spin 1 changes into the state $|-\rangle$ when the photons $ABC$ pass through the cavity 1.
Otherwise, the state of the spin 1 remains unchanged in $|+\rangle$ when the phase information in the two DOFs differs (the hyperentangled state is either one of $\vert \Psi^+_{1}\rangle_P\vert\Phi^-_{1} \rangle_S$ and $\vert \Psi^-_{1}\rangle_P\vert\Phi^+_{1} \rangle_S$). The relationship between
the measurement outcomes of the two spins and the hyperentangled states of the photon system is shown in Table \ref{tab4}. By measuring two-electron spins $1$ and $2$ using the spin basis $\{|+\rangle, |-\rangle\}$ \cite{7,9}, one can determine
the hyperentangled states of the two photons. By far, we have described the scheme of our HGSG using the QD-cavity systems.

\begin{table}[!ht]
\begin{center}
\caption{The relation between the outcomes of the states of the two spins and the obtained final hyperentangled state.}
\begin{tabular}{cccccccc}\hline\hline
  & Spin $1$ and $2$ & & Hyperentangled state \\\hline  & $\vert - \rangle_1\vert -' \rangle_2$ & & $|\Psi^{-}_{1}\rangle_P\otimes|\Phi^{-}_{1}\rangle_S$ \\
& $\vert - \rangle_1\vert +' \rangle_2$ & & $|\Psi^{+}_{1}\rangle_P\otimes|\Phi^{+}_{1}\rangle_S$\\
& $\vert + \rangle_1\vert +' \rangle_2$ & & $|\Psi^{+}_{1}\rangle_P\otimes|\Phi^{-}_{1}\rangle_S$ \\& $\vert + \rangle_1\vert -' \rangle_2$ & & $|\Psi^{-}_{1}\rangle_P\otimes|\Phi^{+}_{1}\rangle_S$\\
\hline\hline
\end{tabular}\label{tab4}
\end{center}
\end{table}

As the HBSG source and the complete and deterministic HGSA are important to quantum communication, it is necessary
to discuss the applications of HGSG and HGSA. Here, we use hyperentanglement
swapping with the present spin-cavity device as an example and describe its application in manipulating multiparticle entanglement (see Fig. 5).
An interesting situation arises when hyperentanglement
swapping is exploited to manipulate multiparticle hyperentanglement.
Hyperentanglement swapping enables multi-parties that are far from each other in
quantum communication to simultaneously share one hyperentangled-state
without directly interacting with each other. First, we assume that three hyperentangled three-photon GHZ states $|\varphi\rangle_{145}$, $|\varphi\rangle_{267}$ and $|\varphi\rangle_{389}$ are prepared in the same state$|\Psi^+_1\rangle_P|\Phi^+_1\rangle_S$ and shared by a central exchanger (Ex) and six users $A, B, C, D, E, F$, and $F$ at different locations in a quantum network (see
Fig. 5). After GHZ hyperentanglement swapping, the whole system which consists of nine photons, becomes
\begin{eqnarray}
&&\frac{1}{8}\;(|\Psi^{+}_{1}\rangle_{123}|\Psi^{+}_{1}\rangle_{456789}+|\Psi^{-}_{1}\rangle_{123}|\Psi^{-}_{1}\rangle_{456789}
+|\Psi^{+}_{2}\rangle_{123}|\Psi^{+}_{2}\rangle_{456789}+|\Psi^{-}_{2}\rangle_{123}|\Psi^{-}_{2}\rangle_{456789}\nonumber\\
&&\;\;+\;\;|\Psi^{+}_{3}\rangle_{123}|\Psi^{+}_{3}\rangle_{456789}+|\Psi^{-}_{3}\rangle_{123}|\Psi^{-}_{3}\rangle_{456789}
+|\Psi^{+}_{4}\rangle_{123}|\Psi^{+}_{4}\rangle_{456789}+|\Psi^{-}_{4}\rangle_{123}|\Psi^{-}_{4}\rangle_{456789})_P\nonumber\\
&&\;\;\otimes\;\;(|\Phi^{+}_{1}\rangle_{123}|\Phi^{+}_{1}\rangle_{456789}+|\Phi^{-}_{1}\rangle_{123}|\Phi^{-}_{1}\rangle_{456789}
+|\Phi^{+}_{2}\rangle_{123}|\Phi^{+}_{2}\rangle_{456789}+|\Phi^{-}_{2}\rangle_{123}|\Phi^{-}_{2}\rangle_{456789}\nonumber\\
&&\;\;+\;\;|\Psi^{+}_{3}\rangle_{123}|\Psi^{+}_{3}\rangle_{456789}+|\Psi^{-}_{3}\rangle_{123}|\Psi^{-}_{3}\rangle_{456789}
+|\Phi^{+}_{4}\rangle_{123}|\Phi^{+}_{4}\rangle_{456789}+|\Phi^{-}_{4}\rangle_{123}|\Phi^{-}_{4}\rangle_{456789})_S,\label{six}
\end{eqnarray}
where
\begin{eqnarray}
|\Psi^{\pm}_{1}\rangle_{456789}=\frac{1}{\sqrt{2}}(|RRRRRR\rangle\pm|LLLLLL\rangle)_{456789},\nonumber\\
|\Psi^{\pm}_{2}\rangle_{456789}=\frac{1}{\sqrt{2}}(|RRRRLL\rangle\pm|LLLLRR\rangle)_{456789},\nonumber\\
|\Psi^{\pm}_{3}\rangle_{456789}=\frac{1}{\sqrt{2}}(|RRLLRR\rangle\pm|LLRRLL\rangle)_{456789},\nonumber\\
|\Psi^{\pm}_{4}\rangle_{456789}=\frac{1}{\sqrt{2}}(|LLRRRR\rangle\pm|RRLLLL\rangle)_{456789},
\end{eqnarray}
and
\begin{eqnarray}
|\Phi^{\pm}_{1}\rangle_{456789}=\frac{1}{\sqrt{2}}(|a_1b_1c_1d_1e_1f_1\rangle\pm|a_2b_2c_2d_2e_2f_2\rangle)_{456789},\nonumber\\
|\Phi^{\pm}_{2}\rangle_{456789}=\frac{1}{\sqrt{2}}(|a_1b_1c_1d_1e_2f_2\rangle\pm|a_2b_2c_2d_2e_1f_1\rangle)_{456789},\nonumber\\
|\Phi^{\pm}_{3}\rangle_{456789}=\frac{1}{\sqrt{2}}(|a_1b_1c_2d_2e_1f_1\rangle\pm|a_2b_2c_1d_1e_2f_2\rangle)_{456789},\nonumber\\
|\Phi^{\pm}_{4}\rangle_{456789}=\frac{1}{\sqrt{2}}(|a_2b_2c_1d_1e_1f_1\rangle\pm|a_1b_1c_2d_2e_2f_2\rangle)_{456789},
\end{eqnarray}
are eight orthogonal GHZ states in the polarization and the spatial-mode DOFs spanning the whole
three-qubit Hilbert space. Here, $x_1$ and $x_2$ ($x=a,b,c,d,e$, and $f$) are different spatial-modes for the photon holed by user $X$ ($X=A,B,C,D,E,$ and $F$).
It is clear from Eq. (\ref{six}) that the
six distant photon belonging to A, B, C, D, E, and F will be
subsequently entangled into one of the 64 GHZ hyperentangled states
according to Ex＊s measurement result. Thus, the generation
and distribution of multiparticle entangled states is
achieved. As an application, this network configuration
can work as a quantum telephone exchanger \cite{quantele}.
By using hyperentangled GHZ states, the transmission capacity of the channels can be twice as effective as normal quantum communication schemes which use  photon-pairs entangled in only one DOF. The quantum network can be extended to N user by using the same principle. This QD-cavity system can also be realized in other physical systems, such atom-cavity system \cite{atomswap} and NV-center-cavity system for similarly relevant levels.

\begin{figure}[!ht]
\begin{center}
\includegraphics[width=8cm,angle=0]{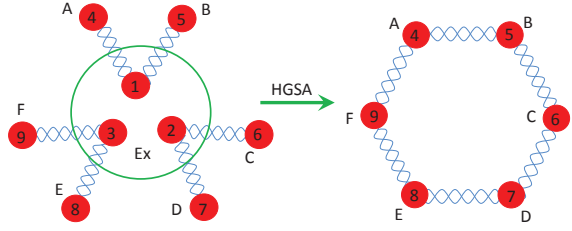}
\caption{Schematic diagram of GHZ hyperentanglement swapping.}
\label{fig.4}
\end{center}
\end{figure}

\section{DISCUSSION AND CONCLUSION}

In this section, we discuss the fidelity of using HBSG and HBSA in a promising system with GaAs-or InAs-based QDs in micropillar microcavities.
The core component of our protocol is the spin-cavity units of which the fidelity and efficiency
are discussed in Ref. \cite{hybrid,7}. In
the present hyperentangled-state analysis scheme, the first spin-cavity unit acts as a SWAP gate,
that directly swaps the initial states in the polarization and spatial-mode DOFs and then
records the relation between the phase information in two DOFs.
\begin{figure}[!ht]
\begin{center}
\includegraphics[width=10cm,angle=0]{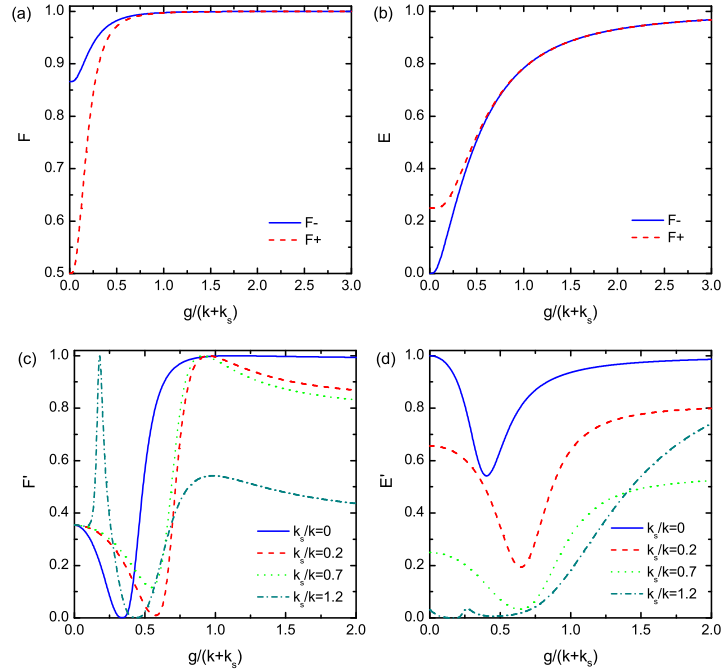}
\caption{Fidelity (a) and efficiency (b) of the double-sided cavity unit via the coupling strength $g/(\kappa+\kappa_s)$ when $|t_0|=|r_h|$. $F_{+}$ and $F_{-}$ represent the fidelity of the present scheme when the outcomes of spin 1 are $|+ \rangle$ and $|-\rangle$, respectively. Fidelity (c) and efficiency (d) of the double-sided cavity unit via the coupling strength $g/(\kappa+\kappa_s)$ and different $\kappa_s/\kappa$.  $F'$ and $E'$ represent the fidelity and efficiency  of the single-sided cavity, respectively.}
\label{fig6}
\end{center}
\end{figure}

To simplify, we consider the case $|t_0(\omega)|=|r_h(\omega)|$, in which the fidelities to
generate (or analyze ) $|\phi^{-}_{AB}\rangle_P\otimes|\phi^{+}_{AB}\rangle_S $, $|\phi^{+}_{AB}\rangle_P\otimes|\psi^{-}_{AB}\rangle_S $, $|\psi^{+}_{AB}\rangle_P\otimes|\psi^{-}_{AB}\rangle_S $, and $|\psi^{-}_{AB}\rangle_P\otimes|\phi^+_{AB}\rangle_S $ can remain in unity,  whereas the fidelity to generate (or analyze) 12 other hyperentangled Bell states are generally less than one, depending on the difference between $|t_0|$ and $|r_0|$. The fidelity (in amplitude) is given by
\begin{eqnarray}
F_{-}=\frac{|r_0^3+t_0^3|}{\sqrt{(r_0^3+t_0^3)^2+3(r_0^2t_0+t_0^2r_0)^2}},\;\;\;\;
F_{+}=\frac{|r_0^3-t_0^3|}{\sqrt{(r_0^3-t_0^3)^2+3(r_0^2t_0-t_0^2r_0)^2}},\label{fidelity}
\end{eqnarray}
and the corresponding efficiencies are
\begin{eqnarray}
E_{-}=(r_0^3+t_0^3)^2+3(r_0^2t_0+t_0^2r_0)^2,\;\;\;\;
E_{+}=(r_0^3-t_0^3)^2+3(r_0^2t_0-t_0^2r_0)^2.\label{effi}
\end{eqnarray}
Here, $F_{+}$ ($E_{+}$) and $F_{-}$ ($E_{-}$) represent the fidelity (efficiency) of the present scheme when the outcomes of the spin 1 states are $|+\rangle$ and $|-\rangle$, respectively.

For the single-sided cavity, the fidelity $F'$ used to measure the phase in the polarization DOF is
\begin{eqnarray}
F'&=&\frac{|r_h'(r_h'^2-3r_0^2)+i(r_0'^2-3r_h'^2)r_0'|}{2\sqrt{2[|r_0'^3|^2+|r_h'^3|^2+3(|r_0'^2r_h'|^2+|r_h'^2r_0'|^2)]}},
\label{fidelity1}
\end{eqnarray}
and the corresponding efficiency $E'$ is
\begin{eqnarray}
E'&=&\frac{|r_0'^3|^2+|r_h'^3|^2+3(|r_0'^2r_h'|^2+|r_h'^2r_0'|^2)}{8}.\label{effi1}
\end{eqnarray}

Fig. \ref{fig6} illustrates numerical calculations of the fidelity and efficiency of the device versus
the coupling strength $g/(\kappa_s+\kappa)$ and different $\kappa_s/\kappa$. A low $\kappa_s/\kappa$ in the strong coupling regime can induce high fidelity and efficiency in both the double-sided and single-sided cavity systems. In the ideal case, where $g/(\kappa +\kappa_s )>1.5$ and
$\kappa_s/\kappa\ll 1$, near-unity fidelity and near-unity efficiency ($F_+=F_-=F'\sim1$, $E>90\%$, and $E'>97\%$ ) can be
simultaneously achieved. Given the cavity loss and $g/(\kappa +\kappa_s )<1.5$, high fidelity still can be achieved but the corresponding efficiency is lower than $0.9$. In the current experiment, a large(7.3 $\mu$m diameter) micropillar cavity can be used to construct the double-sided cavity system. Strong coupling in this cavity has
been observed. The quality factor Q can be improved to $\sim6.5\times10^4$, $\kappa_s/\kappa=0.36$ and the corresponding coupling strength is
$g/(\kappa_s+\kappa)\simeq 0.8$ \cite{24}. As pointed out in
Ref. \cite{13}, a lower $\kappa_s/\kappa$ can be obtained by thinning down the top mirrors of the high-Q
micropillars to achieve effective system parameters: $\kappa_s/\kappa\simeq0.2$ and $Q\sim4.3\times10^4$. In this case, the system now remains in the strong coupling regime with $g/(\kappa+\kappa_s)\simeq0.58$ \cite{thindown}, the fidelities $F_+$ and $F_-$ can reach $0.99$ and the corresponding efficiency $E_{+}=E_{-}=60\%$.
A single-sided spin unit can be constructed by using small micropillar cavities. Effective system parameters: $\kappa_s/\kappa\simeq0.7$,
$g/(\kappa_s+\kappa)\simeq1.0$ can be achieved in strong-coupling
regime with diameter around $1.5 \mu m$ for the In(Ga)As QD-cavity system\cite{13}. A near-unity fidelity can be achievable with $E'= 20\%$. From Figs.\ref{fig6} (c) and \ref{fig6} (d), in the small loss limit ($\kappa_s/\kappa\approx0.2$), a near-unity fidelity $F'\sim0.98$ and efficiency $E'=52\%$ can be achieved at point $g/(\kappa+\kappa_s)=0.88$.
We thus prove that the present scheme can work in the strong-coupling regimes within current technology.

The
fidelities of HGSG operations decreases because of spin decoherence by a factor of $F'=[1+exp(-5t/T)]/2$ \cite{hybrid}, where $T$ is
the electron-spin coherence time ($\sim\mu s$ \cite{5}) and $t$ is the time interval between the different photon spatial-mode states. $5t$ should be considerably shorter than $T$ and $t$ should be longer than $\tau/n_0\sim ns$ \cite{9}, where $\tau$ is the cavity
photon lifetime and $n_0$ is the critical photon number of the
spin-cavity system \cite{52}. The single spin can be measured non-destructively by using photon-spin entanglement. After applying a Hadamard gate on the electron spin, the spin superposition states $|+\rangle$ and $|-\rangle$ can be transformed into the spin states $|\uparrow\rangle$ and $|\downarrow\rangle$, respectively, and the spin can be detected in the $|\uparrow\rangle$ and $|\downarrow\rangle$ basis by measuring the helicity of the transmitted or reflected photon.
As discussed in Ref. \cite{hybrid},
the spin superposition state $|+\rangle$ and $|-\rangle$ can be made from the spin eigenstates by using
nanosecond ESR pulses or picosecond optical pulses \cite{52}.

In summary, we propose a scheme that can be used to not only completely distinguish the 64 hyperentangled GHZ states in both polarization and spatial-mode DOFs but also to produce three-photon hyperentangled GHZ states using the QD-cavity system.
By using numerical calculations, we illustrate that our gate may be feasible for use with the current technologies. In the strong coupling
regime, we expect near-unity fidelity and $> 52\%$ efficiency in both steps of phase measurement. The proposed HGSA and HGSG device
can be applied as a crucial components
in high-capacity, long-distance quantum communication. In this work, we demonstrated how to apply this device in hyperentanglement quantum swapping. With the help of hyperentanglement generation and analysis, the spin-cavity units can not only work in constructing large-scale quantum communication networks but also help in other aspects of quantum
information science and technology.

\section*{ACKNOWLEDGMENTS}

This work was supported by the National Natural Science Foundation
of China under Grant No. 11175094, the National Basic Research
Program of China under Grant Nos. 2009CB929402 and 2011CB9216002,
and the China Postdoctoral Science Foundation under Grant No. 2011M500285.



\begin{thebibliography}{99}
\bibitem{thindown} Here g and $\kappa$ can be controlled independently as g is determined by
the trion oscillator strength and the cavity modal volume, whereas
$\kappa$ by the cavity quality factor only \cite{13}.
\end{thebibliography}
\end{document}